# Speech Emotion Recognition using Support Vector Machine

Manas Jain, Shruthi Narayan, Pratibha Balaji, Bharath K P, Abhijit Bhowmick, Karthik R,
Rajesh Kumar Muthu
School of Electronics Engineering, Vellore Institute of Technology, Vellore.
jainmanas13@gmail.com,shruthi.narayan98@gmail.com, balajipratibha@gmail.com,bharathkp25@gmail.com,
abhijit.bhowmick@vit.ac.in,tkgravikarthik@gmail.com, mrajeshkumar@vit.ac.in.

*Abstract*— **In this project, we aim to classify the speech taken as one of the four emotions namely, sadness, anger, fear and happiness. The samples that have been taken to complete this project are taken from Linguistic Data Consortium (LDC) and UGA database. The important characteristics determined from the samples are energy, pitch, MFCC coefficients, LPCC coefficients and speaker rate. The classifier used to classify these emotional states is Support Vector Machine (SVM) and this is done using two classification strategies: One against All (OAA) and Gender Dependent Classification. Furthermore, a comparative analysis has been conducted between the two and LPCC and MFCC algorithms as well.**

*Keywords— Emotion; datasets; SVM; MFCC; framing; autocorrelation*

## I. INTRODUCTION

Human computer intelligence is an upcoming field of research which aims to make computers learn from experiences and decide how to respond to a particular situation. This has resulted in improved interaction between users and the computer. With the help of certain algorithms and procedures, the computer can be made fit to detect the various characteristics present in the voice sample and deduce the emotion underlying [2]. This emotion detection can be done in two methods, one being speech and the other being image. Since speech is a very important part of communication, it is essential to be able to detect the emotion from it. There are various methods and classifications that are available like K-Nearest Neighbor, Artificial Neural Networks, Hidden Markov Model, Support Vector Machines, and others that have been developed to classify human emotions based on training datasets [2-5].

Feature extraction is the first step that needs to be carried out while detecting emotions. In this project, we have used MFCC and LPC to extract features and then used SVM to train the data sets to identify the emotion or sentiment [5].

SVM is a supervised machine learning technique that is used for classification as well as regression. It attempts to categorize data by finding suitable hyperplanes that can separate data by the highest margin. Based on the training sets, the new values are segregated and analyzed.

In the paper written by Feng Yu etal, they have used support vector machines method to classify basic four types of emotions sadness, happiness, anger and neutral using 721 utterances [2]. In the paper written by Sapra etal, they have used K-nearest Neighbors to classify emotions and used only one technique of MFCC to extract features [4]. In the paper by Utane etal, they have used MFCC to extract features in the speech signal and made a comparative analysis of various other classifiers like Markov model, Gaussian model and support vector machines [5]. El Ayadi etal in his paper has conducted a survey on three important aspects present in speech emotion recognition namely how to design emotional speech corpora, what are the impacts of speech features on performance and classification systems that are present in speech emotion recognition [6]. Kim Samuel etal in their paper has built an application for emotion detection system which operates in real time and uses multi-modal fusion of various features of the speech. [7]. Farouk etal, in his paper have conducted the emotion detection using wavelet analysis and have concluded that wavelet analysis has helped improving the basic features of speech signal as well [8]. Schuller etal, in their paper have used techniques of Hidden markov model to classify emotions. While doing so they have incorporated two varieties in it one that takes global statistics and is classified using Gaussian model and the other where temporal complexity was introduced using several low level instantaneous features[9].Kwon etal, in their paper have used MFCC for feature extraction and have used quadratic discriminant analysis to carry out the speech recognition resulting in decently good results [10].Schuller etal ,in their paper have presented a model firstly by making use of acoustic features after which characteristics like pitch, energy and others are used with respective classifiers to classify human emotions[12].

## II. METHODOLOGY

Speech Emotion Detection aims to identify the emotions of a person based on the input voice sample. This is done by analyzing the input voice signal and carrying out an in depth procedure. First, the speech signal is procured after which features are extracting using MFCC, LPCC which contain the emotional information, voice pattern and coefficients which will be further given as the input to the classification system for further analysis. Our project contains four modules: input speech signal, Feature extraction using MFCC and LPC, Classification based on SVM and the output [3].

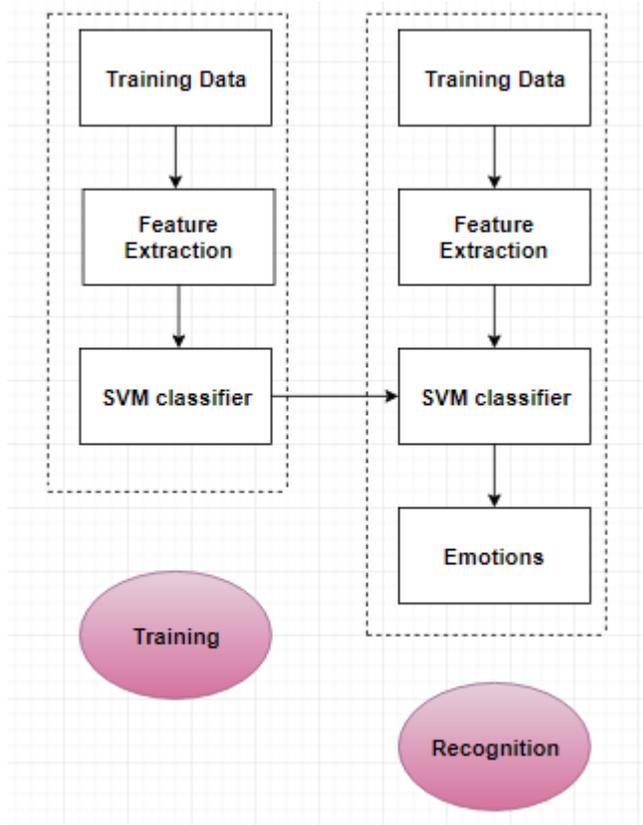

Figure 1. Proposed Model

On extracting features, the following information is collected and they are:

i. Pitch:

The pitch signal is one of the most important features in speech emotion recognition [6]. Pitch frequency is defined as the vibration rate of a vocal. It has all the information about the emotion present because what type of emotion depends on the stress present in the vocal folds and the amount of sub glottal air pressure. Therefore the mean value of pitch present in the samples, variation ranges in the samples and the contours are different in various basic emotional states.

ii. MFCC:

This method of feature extraction has been used widely in feature extraction process. Its calculations are based on the characteristics of human ear, which has a nonlinear frequency unit to make it similar to that of human auditory system. The process of calculating MFCC is shown in Figure 1[9].

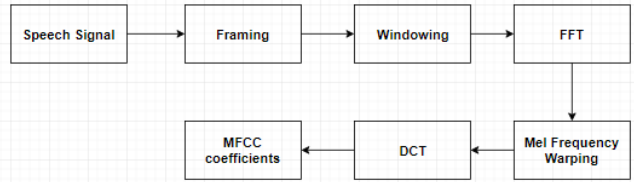

Figure 2. Block diagram of MFCC

iii. LPCC:

It is used for representing an envelope of the input speech signal in compressed form. It uses information from a linear productive model. It analyses the input signal by estimating the formants or enhanced frequency bands. It removes their effects from the signal and estimates the intensity and frequency of the remaining signal. Also this technique is very sensitive to transmission errors.

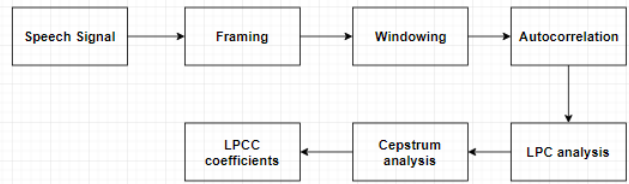

Figure 3. Block diagram of LPCC

iv. Energy:

Energy calculation is one of the most important features of speech analysis [12]. To obtain the complete statistics of all energy features, we use a few short-term functions to extract the value of energy in every speech frame. Using the above information we can obtain the overall statistics of energy for the entire speech sample by calculating the energy such as mean value, maximum value, and variation range present in the voice sample.

v. Speech Rate:

Speech rate shows how fast a human speaks. It has strong correlations with any emotion like happy, sad, fear and angry.

We studied the waveform of the speech signal in time domain; and the speech is calculated using MATLAB function.

### III. SVM ALGORITHM

SVM is a very simple and efficient classifying algorithm which is used for classification and pattern recognition. Support Vector Machines algorithm was introduced by Vladimir Vapnik in 1995.The main aim of this algorithm is to obtain a function that constructs hyper planes or boundaries. These hyper planes are used to separate different categories of input data points. SVM uses binary classification [2].

SVM are systems that use hyper planes in feature space of high dimensions to differentiate values based on a particular specification. Hyper planes are trained with specific algorithms to use statistical learning. SVM method of classification is similar to supervised learning that involves the feature extraction and generates desirable outputs. The advantage of SVM is that it is very easy to train. It can scale high dimensional data better than neural networks [3]. There are typically two types of SVM classifiers, Linear and Non-linear.

Also SVM has kernel functions that can be used in supervising environment. In training phase we are using radial basis kernel function because it limits the training datasets to lie within the specified boundaries. LIBSVM tool is used for SVM classification. During training of signal, feature values of speech signals that are extracted from speech signals are send to LIBSVM with their class labels as Happy, Angry, Sad, and Fear. SVM model is obtained for each emotional state by using their feature values which are extracted.

Once the training model has been prepared with, it is easy to predict the emotional states with testing datasets. Features are extracted from speech signals and with the help of SVM model values generated by training models, the emotions are classified automatically as Happy, Angry, Sad or Fear.

### IV. DATASETS

In this project we have used two datasets, LDC and UGA. The UGA datasets contains the speech samples spoken by students of University of Georgia. It contains 100 samples. We have used 70 samples to train the system and 30 samples for testing. In LDC dataset each speech sample is recorded for one to two seconds with 60ms small segments. The total numbers of voice samples are 100 (70 for training and 30 for testing). For each emotional class there are about 25 utterances spoken by speakers.
Some LDC speech samples are:

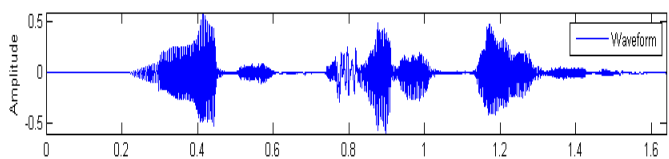

cc_001(m)_hotAnger_4.wav

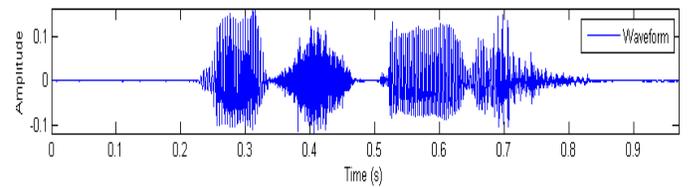

cl_001(m)_happy_active_positive_4.wav

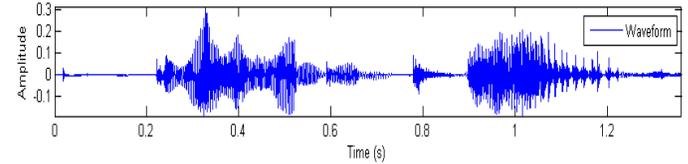

gg_001(f)_sadness_passive_negative_13a.wav

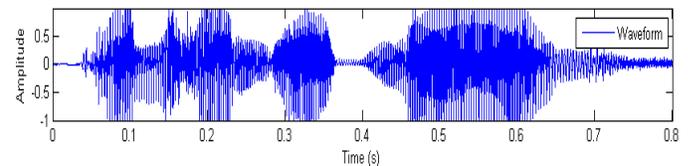

cc_001(m)_hotAnger_4.wav

### V. SIMULATION OUTPUTS AND RESULTS

Waveform and speaking rate of an utterance is shown in Fig. 1 and the features Energy, Pitch and MFCC coefficients and LPCC coefficient are shown in Fig. 3,4,5,6 and 7 respectively.

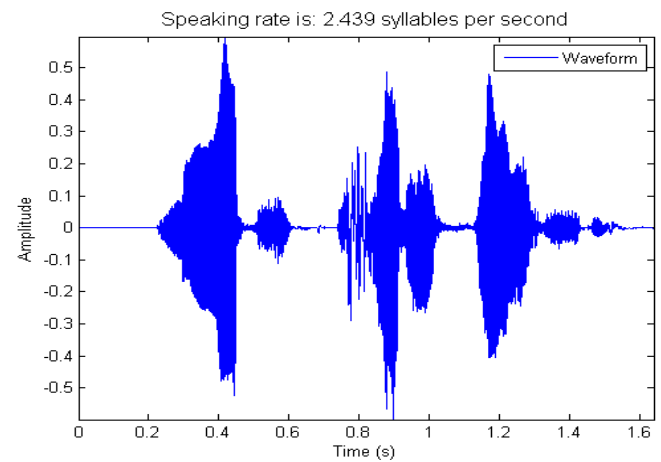

Figure 4. Speaking rate and Waveform

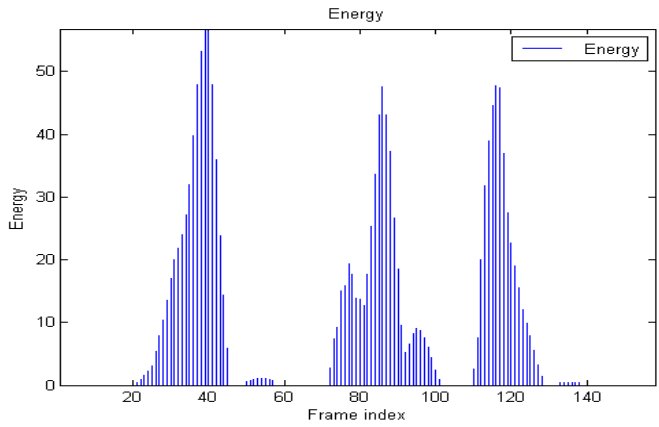

Figure 5. Energy

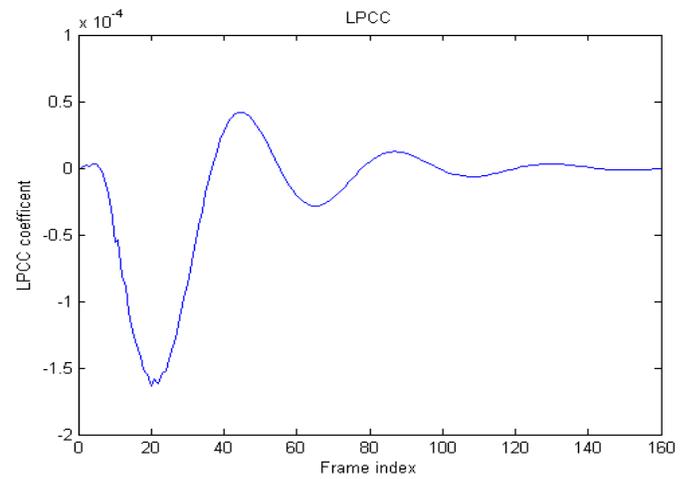

Figure 8. LPCC coefficients

The proposed GUI prototype is shown in Fig. 8 and the final emotion classification result and output for an angry emotion is shown in Fig. 9.

Accuracy of both classifiers (OAA and gender dependent) is shown in table 1 for different emotions whereas table 2 shows the accuracy of both the datasets (LDC and UGA). The overall accuracy using MFCC and LPCC algorithm is also computed and is shown in the table 3.

LDC datasets shows overall accuracy of 90.08% much higher than the accuracy of UGA dataset (65.95 %) as LDC datasets are performed by trained actors and in noiseless environment whereas UGA datasets are performed by students in noisy environment.

The overall accuracy of Gender dependent classifier was found out to be 84.42% higher than the OAA (One against all) SVM classifier (72.785). Graph 1 shows the comparison of accuracies of two classifiers and through that we can conclude that gender dependent classifier shows better accuracy than OAA classifier algorithm. Graph 2 shows the comparison of accuracies of MFCC and LPCC algorithms and it clearly shows MFCC shows better results than LPCC.

The overall accuracy of the system was found out to be 85.085 %.

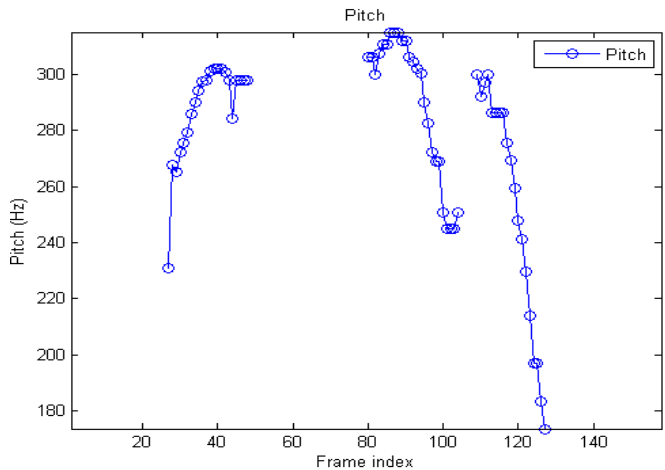

Figure 6. Pitch

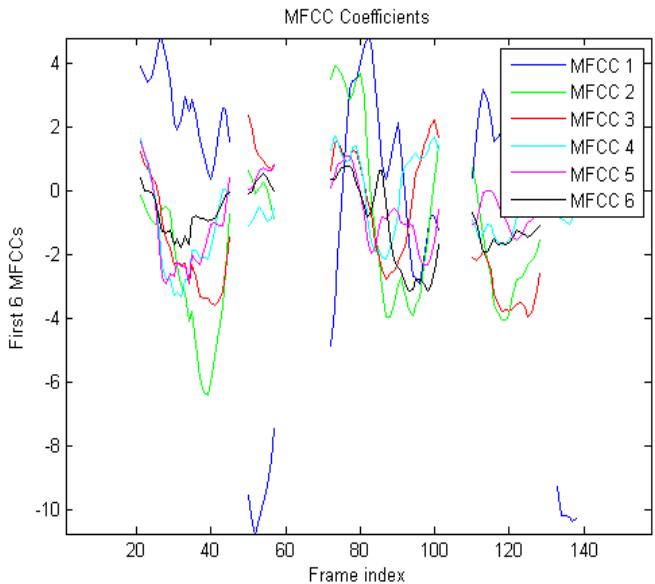

Figure 7. MFCC coefficients

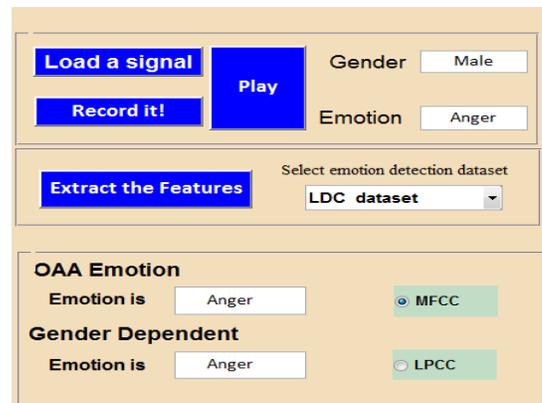

Figure 9. Proposed GUI model

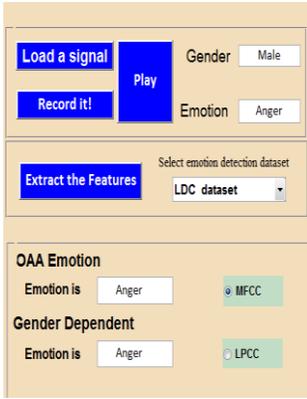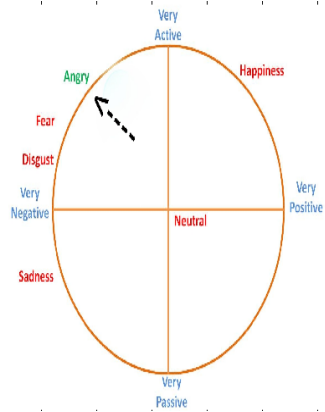

Figure 10. Emotion Classification

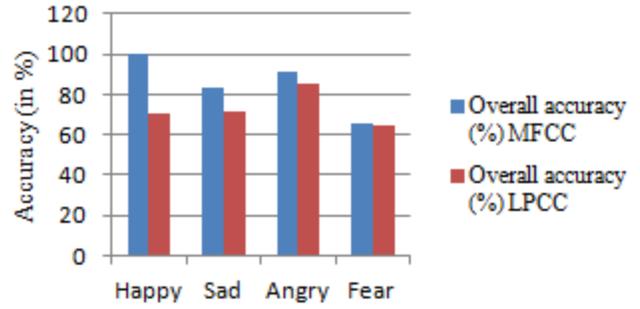

Graph 2. MFCC vs LPCC accuracy comparison

TABLE I: ACCURACY OF BOTH CLASSIFIERS USING MFCC

| Emotions | OAA (%) | Gender dependent (%) |
|---|---|---|
| **Happy** | 94.62 | 98.53 |
| **Sad** | 72.72 | 76.1 |
| **Angry** | 66.66 | 91.66 |
| **Fear** | 57.14 | 71.42 |

TABLE II: ACCURACY FOR BOTH DATASETS USING MFCC

| Emotions | LDC(%) | | UGA(%) | |
|---|---|---|---|---|
| | OAA(%) | Gender dependent(%) | OAA(%) | Gender dependent(%) |
| **Happy** | 98.52 | 99.64 | 42.85 | 60 |
| **Sad** | 63.63 | 83.33 | 54.54 | 66.66 |
| **Angry** | 71.42 | 91.66 | 66.66 | 80 |
| **Fear** | 83.33 | 57.14 | 37.50 | 57.14 |

TABLE III: OVERALL ACCURACY COMPARISON WITH MFCC AND LPCC ALGORITHMS

| Emotions | Overall accuracy (%) | |
|---|---|---|
| | MFCC | LPCC |
| **Happy** | 99.64 | 70.94 |
| **Sad** | 83.33 | 71.32 |
| **Angry** | 91.66 | 85.65 |
| **Fear** | 65.71 | 64.59 |

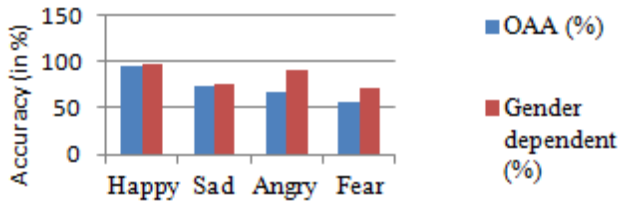

Graph 1: OAA vs Gender dependent classifier

## VI. CONCLUSION AND FUTURE WORKS

Today, the Speech Emotion Recognition has become one of the most important research areas. The type and the number of emotional classes, feature selection, classification algorithm are the important factors of this system. In this project we have used SVM for gaining higher classification accuracy. Two classification strategies (OAA and gender dependent classifier) has been performed and compared. MFCC and LPCC feature values are extracted from speech utterances and emotional states are classified using both the classifiers. On completing the analyses and recording the outputs, we can see that the features extraction using MFCC has garnered a higher degree of accuracy compared to that of LPCC. For training set with acted speakers we obtained recognition rate by **90.08%** with LDC datasets and of **65.97%** with UGA datasets. Since the LDC datasets are recorded in noise less environment and by professional actors, the accuracy of this dataset is very high. The overall accuracy of Gender dependent classifier was found out to be 84.42% higher than the OAA (One against all) SVM classifier (72.785%). We can conclude that gender dependent classifier shows better accuracy than OAA classifier algorithm. The accuracy using LPCC algorithm was found out to be 73.125 % which is less than the overall accuracy using MFCC algorithm which is **85.085 %**.

In our future works, we tend to work on modifying the system by combining another feature values with MFCC and MEDC so as to create the accuracy of system higher. This emotion recognition system is enforced in voice mail application that queues up the calls supported emotions.

### REFERENCES

[1] Rao, K. Sreenivasa, et al. "Emotion recognition from speech." *International Journal of Computer Science and Information Technologies* 3.2 (2012): 3603-3607.

[2] Yu, Feng, et al. "Emotion detection from speech to enrich multimedia content." *Pacific-Rim Conference on Multimedia*. Springer, Berlin, Heidelberg, 2001..

[3] Pfister, Tomas. "Emotion Detection from Speech." *2010*(2010).

[4] Sapra, Ankur, Nikhil Panwar, and Sohan Panwar. "Emotion recognition from speech." *International journal of emerging technology and advanced engineering* 3 (2013): 341-345.

[5] Utane, Akshay S., and S. L. Nalbalwar. "Emotion recognition through Speech." *International Journal of Applied Information Syatems (IJAIS)* (2013): 5-8.

[6] El Ayadi, Moataz, Mohamed S. Kamel, and Fakhri Karray. "Survey on speech emotion recognition: Features, classification schemes, and databases." *Pattern Recognition* 44.3 (2011): 572-587.

[7] Kim, Samuel, et al. "Real-time emotion detection system using speech: Multi-modal fusion of different timescale features." *Multimedia Signal Processing, 2007. MMSP 2007. IEEE 9th Workshop on*. IEEE, 2007.

[8] Farouk, Mohamed Hesham. "Emotion Recognition from Speech." *Application of Wavelets in Speech Processing*. Springer, Cham, 2018. 51-55.

[9] Schuller, Björn, Gerhard Rigoll, and Manfred Lang. "Hidden Markov model-based speech emotion recognition." *Multimedia and Expo, 2003. ICME'03. Proceedings. 2003 International Conference on*. Vol. 1. IEEE, 2003.

[10] Kwon, Oh-Wook, et al. "Emotion recognition by speech signals." *Eighth European Conference on Speech Communication and Technology*. 2003.

[11] Wendemuth, Andreas, et al. "Emotion Recognition from Speech." *Companion Technology*. Springer, Cham, 2017. 409-428.

[12] Schuller, Björn, Gerhard Rigoll, and Manfred Lang. "Speech emotion recognition combining acoustic features and linguistic information in a hybrid support vector machine-belief network architecture." *Acoustics, Speech, and Signal Processing, 2004. Proceedings.(ICASSP'04). IEEE International Conference on*. Vol. 1. IEEE, 2004.

[13] Nwe, Tin Lay, Say Wei Foo, and Liyanage C. De Silva. "Speech emotion recognition using hidden Markov models." *Speech communication* 41.4 (2003): 603-623.

[14] Busso, Carlos, et al. "Iterative feature normalization scheme for automatic emotion detection from speech." *IEEE transactions on affective computing* 4.4 (2013): 386-397.

[15] Sethu, Vidhyasaharan, Eliathamby Ambikairajah, and Julien Epps. "Speaker normalisation for speech-based emotion detection." *Digital Signal Processing, 2007 15th International Conference on*. IEEE, 2007.